\makeatletter
\@ifundefined{@parse@version@dash}{%
\def\@parse@version#1{\@parse@version@0#1}
\def\@parse@version@#1/#2/#3#4#5\@nil{%
\@parse@version@dash#1-#2-#3#4\@nil}
\def\@parse@version@dash#1-#2-#3#4#5\@nil{%
  \if\relax#2\relax\else#1\fi#2#3#4 }
}{}
\makeatother

\documentclass[aps, prb, twocolumn, reprint, superscriptaddress, longbibliography]{revtex4-2}
\usepackage{bm}
\usepackage{color}
\usepackage{graphicx}
\usepackage{adjustbox}
\usepackage{footnote}
\usepackage[utf8x]{inputenc} 

\begin{document}


\title{Re-entrant phenomenon in diffuse ferroelectric, BaSn\textsubscript{0.15}Ti\textsubscript{0.85}O\textsubscript{3} : Local structural insights and  FORC study}

\author{Akash Surampalli}
\affiliation{UGC-DAE Consortium for Scientific Research, University Campus, Khandwa Road, Indore 452001, India.}

\author{Ramon Egli}
\affiliation{Central Institute for Meteorology and Geodynamics (ZAMG), 1190, Vienna, Austria}

\author{Deepak Prajapat}
\affiliation{UGC-DAE Consortium for Scientific Research, University Campus, Khandwa Road, Indore 452001, India.}

\author{Carlo Meneghini}
\affiliation{Dipartimento di Scienze, Universita di Roma Tre, I-00146 Roma, Italy.}


\author{V.Raghavendra Reddy}
\email{varimalla@yahoo.com; vrreddy@csr.res.in}
\affiliation{UGC-DAE Consortium for Scientific Research,  University Campus, Khandwa Road, Indore 452001, India.}

\begin{abstract} 
From the phase diagram as proposed by Lei et.al., \cite{lei2007ferroelectric} a BaSn\textsubscript{0.15}Ti\textsubscript{0.85}O\textsubscript{3} is chosen showing a diffuse phase transition between cubic to rhombohedral (C-R)  near room temperature. Dielectric analysis confirms a phase transition near room temperature ($T$\textsubscript{C} $\approx$ 290 K) and also frequency dispersion in dielectric constant is observed towards low temperature. Polarization and first-order reversal curves (FORC) suggest that the system is in re-entrant phase at low temperatures. Put together, all these electrical characterization results points toward the relaxor behavior in the re-entrant phase. Local probe techniques such as x-ray absorption near edge spectroscopy, Raman and M$\ddot{o}$ssbauer spectroscopy are employed to investigate the local environment changes around this region of low temperature dielectric anomaly. A simple ferroelectric exchange model explaining the low temperature re-entrant behavior is presented from these results. 

\end{abstract}

\pacs{77.80.Jk, 78.70.Dm, 76.80.+y, 74.25.nd}

\keywords{$BaTiO_3$ based relaxors and ferroelectrics, first-order reversal curves, x-ray absorption spectroscopy, Raman spectroscopy, M$\ddot{o}$ssbauer spectroscopy.}

\maketitle 

\section{Introduction}

The ferroelectric (FE) transition with frequency independent broad dielectric peak as a function of temperature is occasionally denoted as diffused phase transition (DPT) in order to distinguish it from the relaxor one \citep{isupov1989some, isupov1993phenomena, smolensky1984ferroelectrics}. Relaxors are usually associated with frequency dependency in their dielectric response and they are made of polar regions of microscopic size (usually in nano-meter range i.e., polar nano-regions (PNR)) showing different electric field response from usual domain like response as observed in FE materials. 

Earliest models proposed for DPT systems are indistinguishable from relaxors, in which the smearing of transition is explained in terms of statistical compositional fluctuations that occur among equivalent crystallographic sites due to doping \citep{smolensky1984ferroelectrics}. Such fluctuations in composition leads to the distribution of local Curie temperatures, $\langle T$\textsubscript{C}$\rangle$, in the system resulting in the broad temperature range of transition. The authors argued that that different local and global symmetries have stringent consequences on their poling-depoling behavior and their low temperature dielectric dispersion \cite{smolensky1984ferroelectrics}. For instance Setter et.al., demonstrated experimentally that introducing disorder to lead-scandium-tantalate (PST, PbSc\textsubscript{x}Ta\textsubscript{1-x}O\textsubscript{3}) makes the ferroelectric transition from sharp to diffuse \cite{setter1980contribution}. 

On the other hand it has been suggested that the mechanism of DPT involved in relaxors differ from the one's exhibiting in FE materials \cite{tsurumi1994mechanism}. Unlike FE, relaxors do not exhibit any phase transition across dielectric maxima and instead the DPT in relaxors is explained as an overlapping phenomenon of volume increase of polar micro regions (which are manifested at high temperatures above dielectric maximum temperature) \& freezing of dipoles \cite{tsurumi1994mechanism}. Comprehending on this idea, further models have been proposed taking analogies with spin-glass models to explain the relaxor behavior \cite{Pirc_Dipolarglass}. Further, experimental evidences of PNR and non-linear dielectric responses as predicted from these models are found \cite{pircnonlinear, tyunina, Liuac}. This led to characterizing relaxors as a separate class from FE. However, with the observation of soft mode like behavior \& domain like responses in the relaxors, the earliest interpretations of relaxors 
have been revived and relaxors are defined as FE with multiple inhomogeneities \cite{fu2009}.  

Barium titanate (BaTiO\textsubscript{3}, BTO) and its modified systems have been well studied for their intriguing physics and applicability \cite{upadhyay, shvartsman2008crossover, shvartsman2009dielectric, upadhyay2014electro}. In particular, the systems in which titanium replaced with an isovalent atom exhibits dielectric and polarization responses that are different from FE systems.  In literature, the compositional phase diagrams of such systems are studied in detail, particularly the phase diagram of BaTi\textsubscript{1-x}Sn\textsubscript{x}O\textsubscript{3} is intensively studied for their excellent properties such as electro-caloric effect \citep{upadhyay2014electro, yasuda, lei2007ferroelectric}. With increasing Sn doping the system exhibits a transformation from FE to relaxors showing diffuse ferroelectric phase transitions for intermediate compositions \cite{lei2007ferroelectric}. 

The system under present investigation, the solid solution BaTi$_{1-x}$Sn$_{x}$O$_3$ for a particular composition (x =0.15) exhibiting DPT, has been a subject to research in the past decade \citep{ansari2019effects,yasuda, upadhyay, shi2015sn, shvartsman2008crossover, xie2012static, deluca2012high}. These past studies are usually focused on the phase segregation that exists in this compound and existence of FE like regions above $\langle T$\textsubscript{C}$\rangle$. Xie et al., observed both static and dynamic like polar regions above $\langle T$\textsubscript{C}$\rangle$, transforming to a long range FE below  $\langle T$\textsubscript{C}$\rangle$ \cite{xie2012static}. Similar observations are made by Shvartsman et al., in which they argued the phase transition occurring in BaTi$_{0.85}$Sn$_{0.15}$O$_3$ is different from conventional FE transition \cite{shvartsman2008crossover}. Precursor polar like regions that are responsible for transitions are seen above dielectric maxima ($T$\textsubscript{m}) and the  bulk transformation to FE is completed at a much lower temperature (\textit{T} $<$ $T$\textsubscript{m}). However, in the later work published by Sanjay et al., of the similar composition, it is conjectured that towards low temperature relaxor phase sets in co-existing with FE \cite{upadhyay}. This is interpreted from the changes in polarization hysteresis curves in the electric field cooled conditions and re-entrant like behavior in remanent polarization variation with temperature. 

There is no further consensus available on the above mentioned aspects and in general the re-entrant phenomenon is very meagerly explored in the context of FE materials as compared to their magnetic counterparts \cite{senoussi1988magnetic, sato2001spin, dho2002reentrant}. Re-entrant phases are usually associated with disordered phase that is followed by an ordered phase on lowering the temperature. Such disorder is likely to occur by breaking the ordered state into a disordered state or by having different phases co-exist at different scales \cite{bokov2016reentrant}. Re-entrant phenomenon in ferroelectrics are distinguished in materials which reveals atypical sequence of phase transitions viz., paraelectric-ferroelectric-relaxor on lowering temperature. These phase transitions are identified by decrease in polarization, dispersion in dielectric constant and an associated loss peak, aging at low temperatures \cite{bharadwaja2011critical, li2013re, bharadwaja2012reentrant, fang2018re}. Investigating re-entrant transitions in ferroelectrics is especially challenging since the properties that are associated with them can be explained with external process i.e., domain wall \& domain responses . Hence, the idea that re-entrant phases in ferroelectrics are associated with a thermodynamic phase transition is received with skepticism \cite{li2013re}.

In view of this, in the present work we have taken the system BaSn\textsubscript{0.15}Ti\textsubscript{0.85}O\textsubscript{3} in which the properties associated with re-entrant behavior are observed \cite{upadhyay}. To understand the polarization evolution in the system, we have employed first-order reversal curves (FORC) to reveal subtle changes that are responsible for the re-entrant behavior or to find out if such behavior exists. Along with it, x-ray absorption spectroscopy (XAS), M$\ddot{o}$ssbauer and Raman spectroscopy techniques are further employed to understand the structural changes occurring across this re-entrant temperature. It may be noted that combined use of such microscopic and macroscopic experimental techniques has revealed structural modifications associated to the relaxor behavior in BTO based compounds in our recent work \cite{surampalli2019evidence}.

  
\maketitle \section{Experimental details}

Polycrystalline 15\% Sn doped BaTiO\textsubscript{3} (BaSn\textsubscript{0.15}Ti\textsubscript{0.85}O\textsubscript{3}) is prepared with conventional solid-state sintering method starting from high purity ($\geq$99.9\%)  oxide and carbonate precursors. X-ray diffration measurements are carried out with Brucker D8-Discover system equipped with Cu K\textsubscript{$\alpha$} radiation and LynxEye detector. Dielectric measurements were performed using precision LCR meter E4980 and in-house developed sample holder with 1 V ac signal at different frequencies in the temperature range 50-300 K, while cooling the sample. Temperature dependent P-E hysteresis loops and first order reversal curves (FORC) were measured using Radiant Precision Premier II ferroelectric loop tracer using liquid nitrogen as cryogen in the temperature range of 80-350 K and placing the sample in Delta-9023 chamber. 
Temperature dependent Raman spectra were recorded with LabRam-HR800 in back scattering geometry equipped with the diode laser ($\lambda$=473 nm) as excitation source in the temperature range 80-325 K using Linkam cryostat. 
%
Temperature-dependent Ti K edge x-ray absorption spectra (XAS) data were collected in transmission geometry at ELETTRA (Trieste, Italy) 11.1R beamline \citep{Elettra}. Absorption spectra were measured in transmission geometry using Ti foil as a reference sample for energy calibration. The raw data were treated using standard procedures for normalization and extraction of XAFS signal \cite{Meneghini}. Temperature dependent \textsuperscript{119}Sn M$\ddot{o}$ssbauer measurements are carried out in transmission mode using a standard PC-based M$\ddot{o}$ssbauer spectrometer equipped with a WisEl velocity drive in constant acceleration mode and placing the sample in a Janis CCR system. The spectrometer is calibrated with natural iron and the reported isomer shift values are with respect to SnO\textsubscript{2}.  

\begin{figure}[b]
  \centering
  \includegraphics[width=8cm, scale=2]{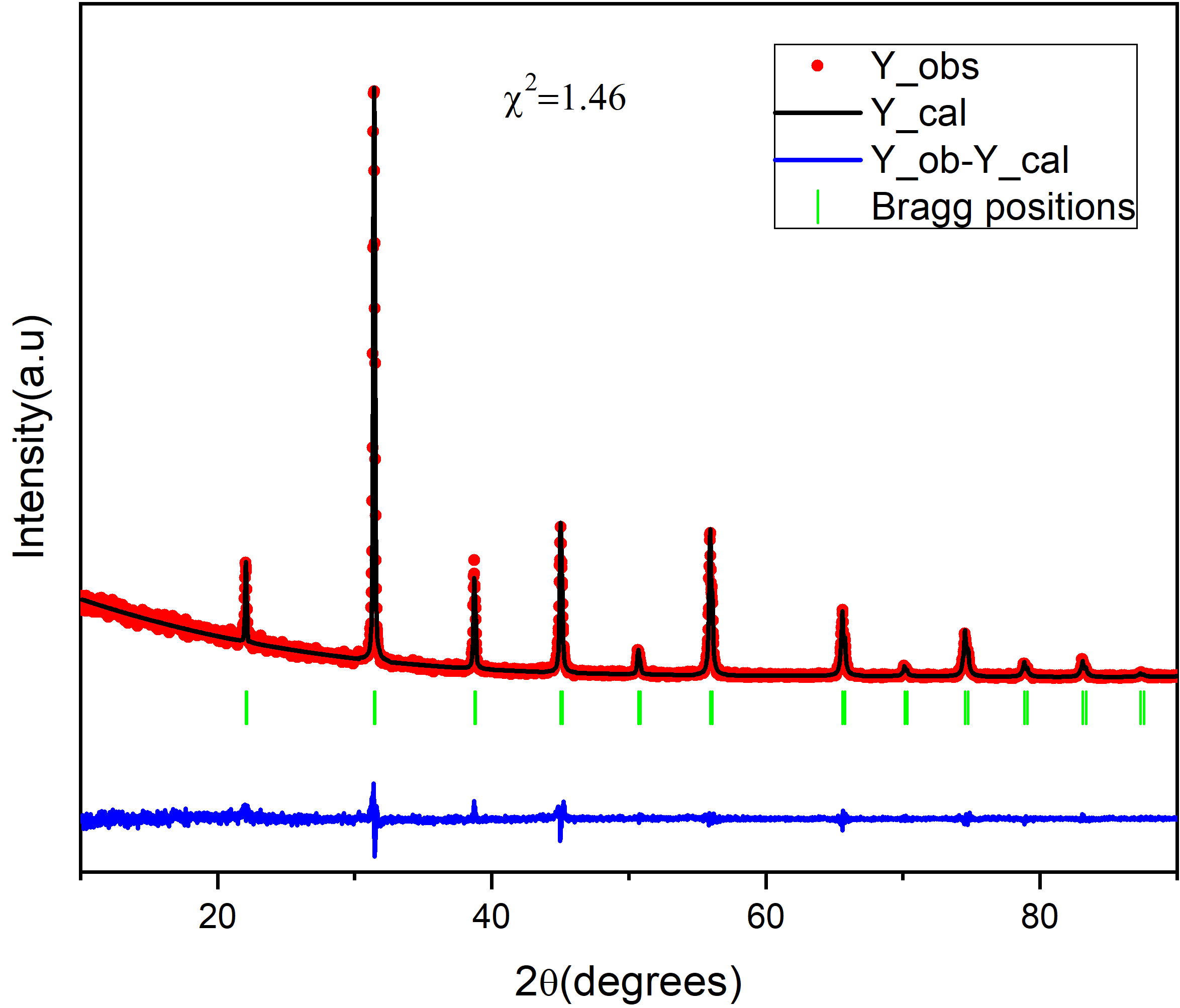}
  \caption{Room temperature x-ray diffraction pattern of BaSn\textsubscript{0.15}Ti\textsubscript{0.85}O\textsubscript{3} measured with $\lambda=0.154$ nm radiation.  The experimental data is profile fitted with FullProf\cite{rodriguez1993recent} program.}
  \label{fig:xrd}
\end{figure}

\maketitle \section{Results and discussions}

The X-ray powder diffraction data obtained at room temperature (shown in Fig.~\ref{fig:xrd}) reveals a Pm$\overline{3}$m cubic symmetry consistent with the literature \cite{lei2007ferroelectric}. The $\chi$\textsuperscript{2} value is found to be well within the accepted range. The XRD data indicates the phase purity of the studied sample and the results of all the measurements are discussed in the following sections. 

\begin{figure*}[htbp]
  \centering
  \includegraphics[height= 8cm, width=\textwidth, keepaspectratio]{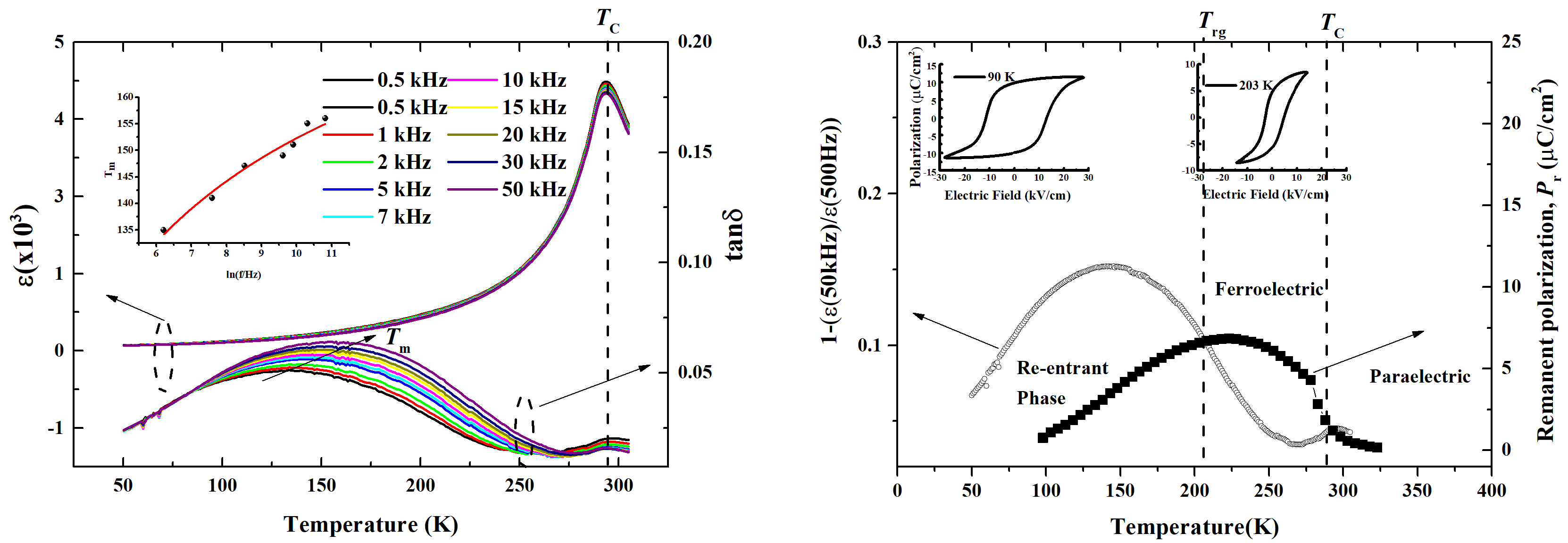}
  \caption{(a) Temperature dependent dielectric constant ($\epsilon$) and tan$\delta$ of BaSn\textsubscript{0.15}Ti\textsubscript{0.85}O\textsubscript{3} at different frequencies. The inset shows the Vogel-Fulcher fit to the variation of $T$\textsubscript{m} as a function of frequency. (b) Variation of 1-($\epsilon$(50 kHz)/$\epsilon$(500 Hz)) as a function of temperature of BaSn\textsubscript{0.15}Ti\textsubscript{0.85}O\textsubscript{3} and remanent polarization, $P$\textsubscript{r} vs temperature as obtained from PUND measurements, taken at a electric field of around 8 kV/cm. The inset shows the P-E loops  at the indicated temperatures.}
  \label{fig:Dielec_PE}
\end{figure*}

\begin{figure*}[htbp]
  \centering
  \includegraphics[height= 8cm, width=\textwidth, keepaspectratio]{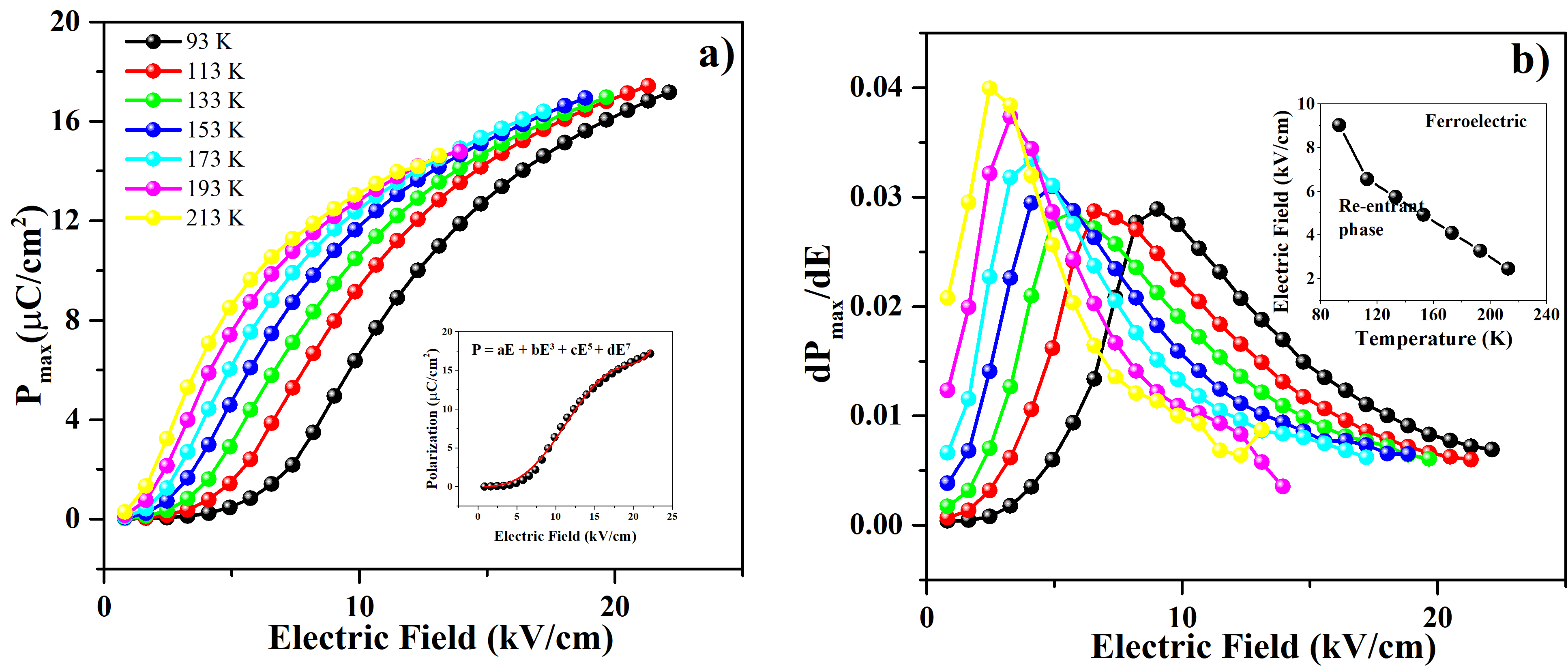}
  \caption{(a) Maximum electric polarization ($P$\textsubscript{max}) plotted as a function of electric field, taken from the temperature dependent P-E data. The inset shows polynomial fitting for 80 K as adopted in literature \cite{prosandeev2013field}. (b) Its derivative \& the inset shows the electric field corresponding to derivative  maxima against the temperature.}
  \label{fig:pmax}
\end{figure*}

\subsection{Temperature dependent dielectric and electric polarization studies}

The temperature dependent dielectric constant ($\epsilon$) and dielectric loss (tan$\delta$) are shown in Fig.~\ref{fig:Dielec_PE}.  On cooling  it can be seen from the Fig.~\ref{fig:Dielec_PE}(a) that at around 290 K ($T$\textsubscript{C}), a broad peak in dielectric constant is seen, a characteristic of DPT, as well as a peak in dielectric loss near the same temperature. This is a feature reminiscent of paraelectric-ferroelectric transition ($T$\textsubscript{C}). Upon further lowering the temperature, frequency dispersion in dielectric constant becomes evident and a steady increase in loss is seen with frequency dependent dispersive peak. The loss peaks roughly follows Vogel-Fulcher law indicating a possible existence of dipolar freezing dipole dynamics in the system. The frequency dispersion in $\epsilon$ represented as 1-$\frac{\epsilon\textsubscript{50 kHz}}{\epsilon\textsubscript{500 kHz}}$ shows a behavior similar to tan$\delta$ as shown in Fig.~\ref{fig:Dielec_PE}(b). The frequency dispersion start to appear above  $T$\textsubscript{C}, showing a local maxima around $T$\textsubscript{C}. The relaxation around dielectric maxima had been already discussed by shvartsman et.al, that relaxing polar entities manifested at high temperatures percolating to ferroelectric state are responsible for this dispersion around dielectric maxima \cite{shvartsman2008crossover}. This FE transition is accompanied with increasing dispersion on further lowering the temperature, peaking at a temperature and starts to vanish below that. In order to investigate such behavior at low temperatures, electric polarization versus electric field (P-E) loop evolution with temperature and remanent polarization obtained from positive up and negative down (PUND) measurements are performed and are also shown in Fig.~\ref{fig:Dielec_PE}(b) comparing with the trend in dielectric dispersion in the system.

Apart from the dispersion in dielectric constant, remanent polarization obtained from PUND data is seen to be decreasing towards lower temperatures (Fig.~\ref{fig:Dielec_PE}(b)), peaking at temperature $T$\textsubscript{rg} deviating from the trend in normal ferroelectrics. It is also observed that this value of $T$\textsubscript{rg} is electric field dependent (not shown here) and at higher fields the trend follows as such of normal ferroelectrics. These results suggest that the system transforms into metastable FE state at low temperature and can be stabilized into a FE state on the application of high electric fields. Field cooling protocols have been adopted in earlier studies to probe the metastable nature of this system \cite{upadhyay}. The coercive switching fields are also seen to be increasing below $T$\textsubscript{rg} as seen from hysteresis curves in Fig.~\ref{fig:Dielec_PE}. Our results are strikingly similar to the dielectric and electric polarization studies in (Ba\textsubscript{0.925}Bi\textsubscript{0.05})(Ti\textsubscript{1−x}Sn\textsubscript{x})O\textsubscript{3} system, where a re-entrant transition from ferroelectric to glass like state is observed \cite{fang2018re} from transmission electron microscopic measurements. Adopting similar nomenclature used in the literature, this characteristic temperature T\textsubscript{rg} is taken as re-entrant temperature. However, with further analysis we establish the true nature associated with this temperature. 

Further, the electric polarization trend with electric fields obtained from hysteresis curves taken at different temperatures are studied to realize the nature of this re-entrant state and the data is shown in Fig.~\ref{fig:pmax}. The trend obtained follows the polynomial that is adopted for relaxor studies \cite{prosandeev2013field} as shown in the inset of Fig.~\ref{fig:pmax}(a). And the derivative as also shown in Fig.~\ref{fig:pmax}(b), which corresponds to static dielectric constant, is plotted against electric field. The derivative, which shows a maxima corresponds to the transformation from the relaxor ferroelectric state to long-range ferroelectric \cite{prosandeev2013field}. It is also conjectured by Sanjay et al., \cite{upadhyay} that this system co-exists in ferroelectric \& relaxor state towards low temperatures. The results of electric polarization with electric field \& dielectric dispersion of Vogel-Fulcher like points towards the relaxor like signatures that exist in the exist. These results led us to believe that the metastable nature of this system is given by the relaxor like properties that are observed. To further understand the evolution of electric polarization in such systems first order reversal curves are measured and subsequently analyzed as discussed in the following section.

\subsection{First Order Reversal Curves}

\begin{figure*}[htbp]
  \centering
  \includegraphics[height= 15cm, width=\textwidth, keepaspectratio]{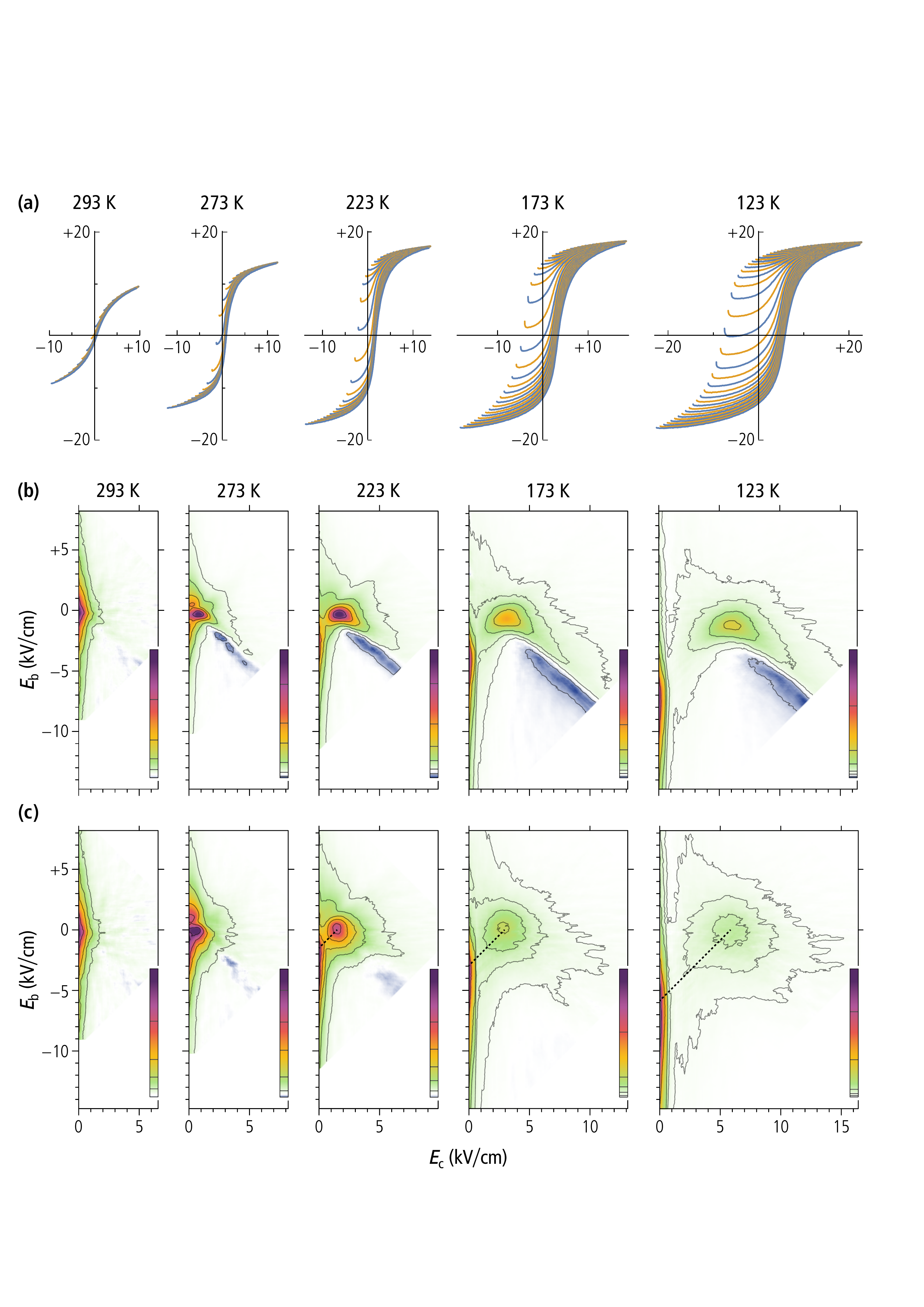}
  \caption{ FORC measurements at 293, 273, 223, 173, and 123 K (every fifth curve is shown for clarity, electric field in kV/cm and polarization in $\mu$C/cm$^2$) . (b) FORC diagrams corresponding to the measurements in (a). Contours enclose polarizations corresponding to the 15, 40, 60, and 80 \% quantiles of the total integral of the FORC function. (c) FORC diagrams corresponding to the measurements in (a), expressed as a function of the total internal field $E$\textsubscript{i} = $E$+$\alpha P$/$\epsilon$\textsubscript{0}, instead of the applied field $E$, with coefficients $\alpha$ given in Table 1. The dashed lines connect the peaks of the thermal relaxation feature on the left margin with the central maximum along the trace of a single curve.}
  \label{fig:forc}
\end{figure*} 

\begin{figure}[b]
  \centering
  \includegraphics[width=8cm]{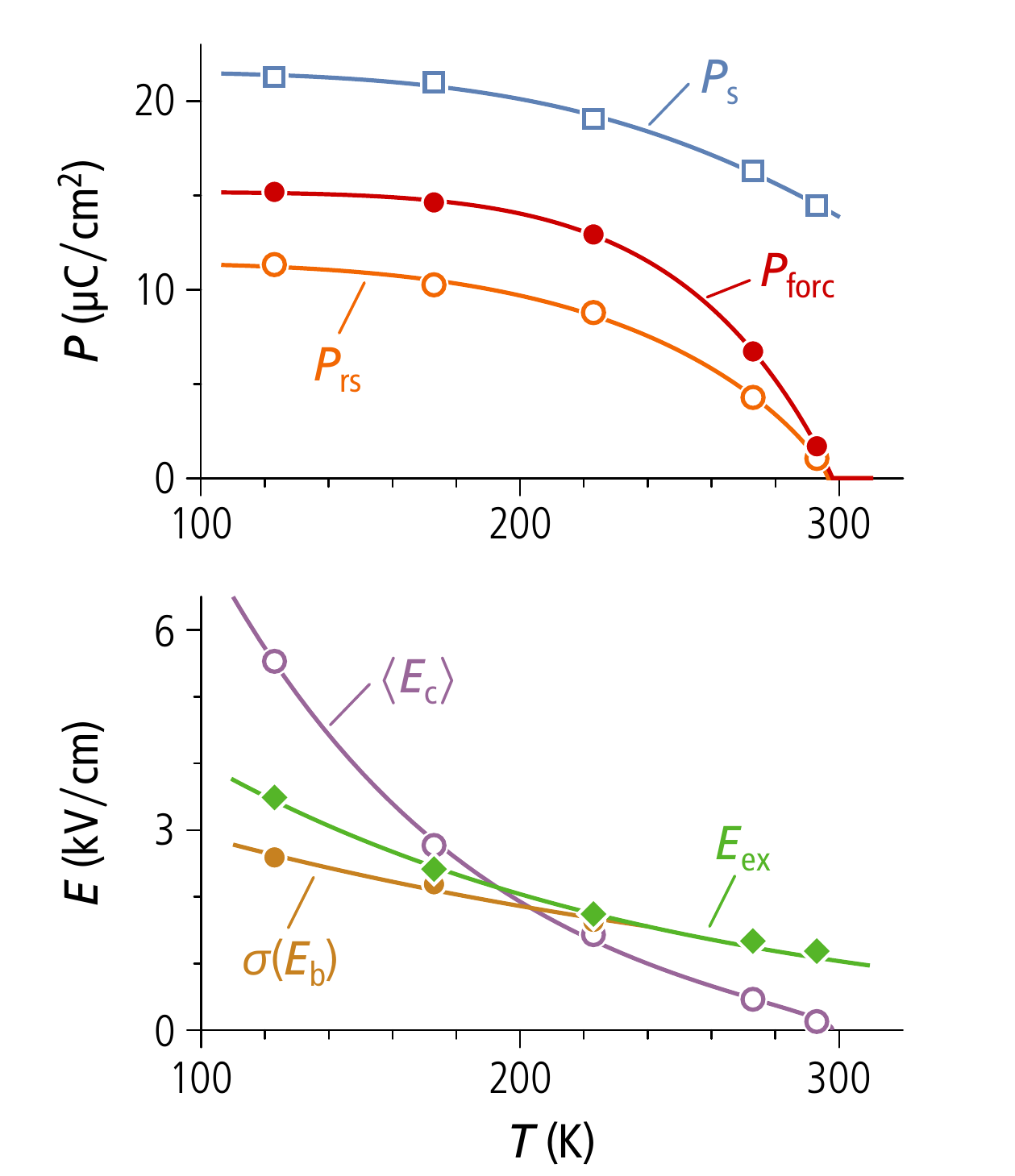}
  \caption{ (a) Saturation polarization $P$\textsubscript{s} (squares), remanent saturation polarization $P$\textsubscript{rs} and total FORC polarization $P$\textsubscript{forc} (dots) obtained from the desheared FORC measurements shown in \ref{fig:forc}. Lines are guides to the eye. (b) Temperature dependencies of the exchange field $E$\textsubscript{ex} (diamonds), the median switching field $\langle E$\textsubscript{c}$\rangle$ (circles), defined as the median of the coercivity distribution represented by horizontal profiles of the FORC functions in \ref{fig:forc}c at $E$\textsubscript{b}=0 , and the standard deviation of the random bias field $E$\textsubscript{b} (dots), obtained from vertical profiles of the FORC function in \ref{fig:forc}c across the central maximum. Lines are guides to the eye.}
  \label{fig:summary}
\end{figure}

First-order reversal curves $\textit{P}(\textit{E}_{r},\textit{E})$ \cite{franco} have been measured at selected temperatures, starting from room temperature (Fig.\ref{fig:forc}a). Each curve is preceded by a preparation step, where the applied field is
decreased from positive saturation to the so-called reversal field \textit{E}\textsubscript{r}, and consists in measurements of \textit{P} while the applied field \textit{E} is increased from \textit{E}\textsubscript{r} to positive saturation. The FORC function $\rho$=$\partial$$^2$$P$/($\partial$\textit{E}\textsubscript{r}$\partial$\textit{E}) has been calculated with the VARIFORC protocol using local polynomial fits \cite{egli2013variforc}, and plotted in modified coordinates \textit{E}\textsubscript{c}$=$(\textit{E}-\textit{E}\textsubscript{r})/2 and \textit{E}\textsubscript{b}$=$(\textit{E}+\textit{E}\textsubscript{r})/2 (Fig.\ref{fig:forc}b). In Preisach theory \cite{franco}, \textit{E}\textsubscript{c} and \textit{E}\textsubscript{b} are the coercive and bias fields of elemental rectangular hysteresis loops, called hysterons, and the FORC function is the joined probability density of these fields, which is required to generate the measured curves. The room temperature ($\sim$293 K) FORC function is concentrated along the left margin (\textit{E}\textsubscript{c}=0 ) and has a large vertical spread with a single maximum close to the origin. This signature is typical of a system containing regions whose remanent polarization \textit{P}(\textit{E}=0) decays in a time comparable with that of measurements, due to thermal relaxation \cite{lanci2018forward}. The hysteresis becomes completely closed above $T$\textsubscript{C}, yielding $\rho$=0. The approach to $T$\textsubscript{C} is clearly visible on the temperature dependence of the remanent saturation polarization $P$\textsubscript{rs}=\textit{P}(\textit{E}\textsubscript{r}, \textit{E}=0) and of the total irreversible polarization $P$\textsubscript{forc} obtained by integrating the FORC function over $E$\textsubscript{c} and $E$\textsubscript{b} (Table 1). Both quantities disappear at $T$\textsubscript{C} $\approx$297 K (Fig.\ref{fig:summary}a), in correspondence with the peak of the dielectric constant (Fig. 2). For comparison, the saturation polarization $P$\textsubscript{s} displays only a moderate decrease in proximity of $T$\textsubscript{C}.

Isolated regions undergoing thermal relaxation near \textit{$T$\textsubscript{C}} produce FORC functions characterized by a vertical ridge at $E$\textsubscript{c}=0, which extends mainly over $E$\textsubscript{b}$\lesssim$ 0, since a negative reversal field is required to switch these regions \cite{lanci2018forward}. In the formalism of the Preisach theory, the almost symmetric ridge observed at room temperature requires a random bias field $E$\textsubscript{b} with a broad probability distribution, which acts as a thermal fluctuation field with spatial and time correlation \cite{berkov2004micromagnetic}. The existence of a strong random coupling between relaxing regions is also suggested by the amplitude-dependent, frequency independent maximum of tan$\delta$ in measurements of the dielectric constant around $\sim$290 K (Fig. 2), which contrasts with the amplitude-independent and frequency-independent behavior expected from thermal relaxations in non-interacting systems \cite{jonsson2004superparamagnetism}.

A second FORC feature emerges at lower temperatures. This feature consists of a second peak, hereafter called the central peak, which forms near the origin (Fig.\ref{fig:forc}b, 273 K), and moves towards higher values of E$_c$ as the temperature decreases. A similar feature, consisting of a horizontal ridge along $E$\textsubscript{b}=0, is produced by the switching of isolated regions that become progressively blocked \cite{lanci2018forward}; however, the central maximum seen here is characterized by oval contour lines, which, like in the case of the vertical ridge caused by thermal relaxation, denote the presence of random interactions \cite{carvallo2005experimental}. In the context of Preisach theory, such interactions are represented by a local internal field whose probability distribution density is given by vertical profiles through the FORC function \cite{egli2006theoretical, franco}. The central peak is offset towards negative $E$\textsubscript{b} values comprised between $\sim$0.2 kV/cm (273 K) and $\sim$1.2 kV/cm (123 K), which can be understood as a positive mean internal field that promote switching to positive saturation in external fields that are smaller than those required without this internal field. A pair of positive and negative ridges parallel to the descending diagonal departs from the central peak, while the vertical ridge near the left edge of the FORC diagram is shifted downwards. Similar signatures have been observed in polycrystalline exchange-spring magnets consisting in a magnetically hard/soft bilayer \cite{davies2005anisotropy}, and in polycrystalline iron thin films \cite{cao2015hysteresis}. The development of the diagonal ridge pair has also been associated with the formation of long-range ferromagnetic order between ferromagnetic clusters embedded in a non-ferromagnetic matrix \cite{davies2005magnetization}.

Ferroelectric exchange coupling can be represented by a so-called exchange electric field $E$\textsubscript{ex}=$\alpha P$ that is proportional to the bulk polarization through a positive coefficient $\alpha$. In this case, switching of ferroelectric regions depends on the total field $E$\textsubscript{tot} = $E$+$\alpha P$ resulting from the overlap of $E$\textsubscript{ex} with the field \textit{E} applied during FORC measurements. Electric polarization curves expressed as a function of $E$\textsubscript{tot}, instead of the applied field, represent the intrinsic response of the switching regions without any coupling between them. This operation is called deshearing \cite{franco}. If FE exchange is representable by the above model, a value of $\alpha$ can be found for which the diagonal ridge pair is eliminated from the desheared FORC diagram, while the central maximum moves to the $E$\textsubscript{b}=0 position expected for unbiased switching regions. This is indeed the case (Fig.\ref{fig:forc}c), with increasingly large values of $\alpha$ required at lower temperatures (Table 1). Hysteresis deshearing with $\alpha>$0 causes a decrease of $P$\textsubscript{rs}, whereby $P$\textsubscript{rs}/$P$\textsubscript{s} $\approx$ 0.5 is obtained from desheared $P$\textsubscript{rs} values at low temperatures. This result is expected for isolated and completely blocked regions with uniaxial ferroelectric anisotropy.

The amplitude of the central maximum in desheared FORC diagrams decreases with decreasing temperature; however, this is just due to its broadening, since $P$\textsubscript{forc} increases monotonically as the sample is cooled down (Fig.\ref{fig:summary}). The median $\langle E$\textsubscript{c}$\rangle$ of the distribution obtained from horizontal profiles of the FORC function (Table 1) is a measure for the mean switching field. It is characterized by a strong temperature dependence and vanishes near $T$\textsubscript{C} because of thermal relaxations. The vertical extension of the desheared central maximum is a measure for the strength of random interactions between switching regions, expressed for instance by the standard deviation $\sigma$(E$_b$ ) of a probability density function identified by vertical profiles across the central maximum. Estimates of $\sigma$($E$\textsubscript{b}) have been obtained for \textit{T}$\le$223 K (Table 1); above this temperature, the central maximum overlaps with the vertical ridge caused by thermal relaxation yielding unreliable results. The similarity between magnitude and temperature dependence of $E$\textsubscript{ex} and $\sigma$($E$\textsubscript{b}) (Fig.\ref{fig:summary}b) points to a common origin of these two quantities, which can be understood as the mean and the standard deviation of a positive random variate describing the intensity of local ferroelectric exchanges. As in the case of thermal relaxation near $\sim$220 K, the presence of an interaction field is confirmed by the frequency-dependent amplitude maximum of tan$\delta$ in measurements of the dielectric constant in the re-entrant spin glass range (Fig.\ref{fig:Dielec_PE}a).

Further, the peak of the thermal relaxation feature along the left edge of the desheared FORC diagrams is connected to the central maximum by a $45^{\circ}$ line (dotted in Fig.\ref{fig:forc}c), which means that curves with most switching are also those affected by the largest relaxation effects associated with the hook-shaped initial part (Fig. \ref{fig:forc}a). The polarization decrease at the beginning of FORCs can be explained by the thermally activated switching of positively saturated regions in proximity of the negative reversal field \cite{lanci2018forward}. The same regions are switched back to positive saturation in correspondence of the central maximum of the FORC diagram.

Further, to demonstrate the structural changes that are associated with the electric polarization evolution, spectroscopic techniques viz., Mossbauer, Raman \& XANES are employed and are discussed in the following sections.

\begin{table*}[htbp]
\caption{\label{arttype} Parameters derived from FORC measurements: $P$\textsubscript{s} — saturation polarization, $P$\textsubscript{rs} — remanent saturation polarization, $P$\textsubscript{rsd} — remanent saturation polarization obtained from desheared measurements, $P$\textsubscript{forc} — integral of the FORC function over the measured domain, $\langle E$\textsubscript{c}$\rangle$ — median switching field calculated from horizontal profiles of the desheared FORC function at $E$\textsubscript{b}=0 and over $E$\textsubscript{c}$\geq$ 0.6 kV/cm (to exclude contributions from the vertical ridge), $\alpha$ — deshearing coefficient, $\sigma$($E$\textsubscript{b}) — standard deviation of the distribution of bias fields obtained from vertical profiles of the desheared FORC function through the central maximum, $E$\textsubscript{ex}=$\alpha P$ — mean exchange field.}
\begin{ruledtabular}
\begin{tabular}{cccccccccc}
\textit{T}  & $P$\textsubscript{s}  & $P$\textsubscript{rs}   & $P$\textsubscript{rsd}   & $P$\textsubscript{forc} & $P$\textsubscript{rsd}/$P$\textsubscript{forc} & $\langle E$\textsubscript{c}$\rangle$ & $\alpha$  &$\sigma$($E$\textsubscript{b}) & $E$\textsubscript{ex}    \\	
(K) &$\mu$C/cm$^2$ &$\mu$C/cm$^2$ &$\mu$C/cm$^2$ &$\mu$C/cm$^2$ &- &kV/cm &kVm/C &kV/cm &kV/cm \\
\hline
293		&14.5	&1.26	&1.05 &1.69 &0.073 &0.43 &$\sim$ 820 & $^{a}$   &1.19 \\
273		&16.3	&6.16	&4.28 &6.72 &0.26 &0.76 &$\sim$ 820  & $^{a}$    &1.34 \\
223		&19.1	&10.5	&8.78 &12.9 &0.46 &1.73 & 820        & 1.62 &1.56 \\
173		&21.0	&12.2	&10.3 &14.6 &0.49 &3.07 &1150        & 2.19 &2.41 \\
123		&21.3	&13 	&11.3 &15.2 &0.53 &5.83 &1640        & 2.59 &3.49 \\
\footnote {Not determined because of the overlap with the vertical ridge.}
\end{tabular}
\end{ruledtabular}
\label{tab:parameters}
\end{table*}

\subsection{Raman spectroscopy investigation}

\begin{figure}[htbp]
  \centering
  \includegraphics[width=8cm]{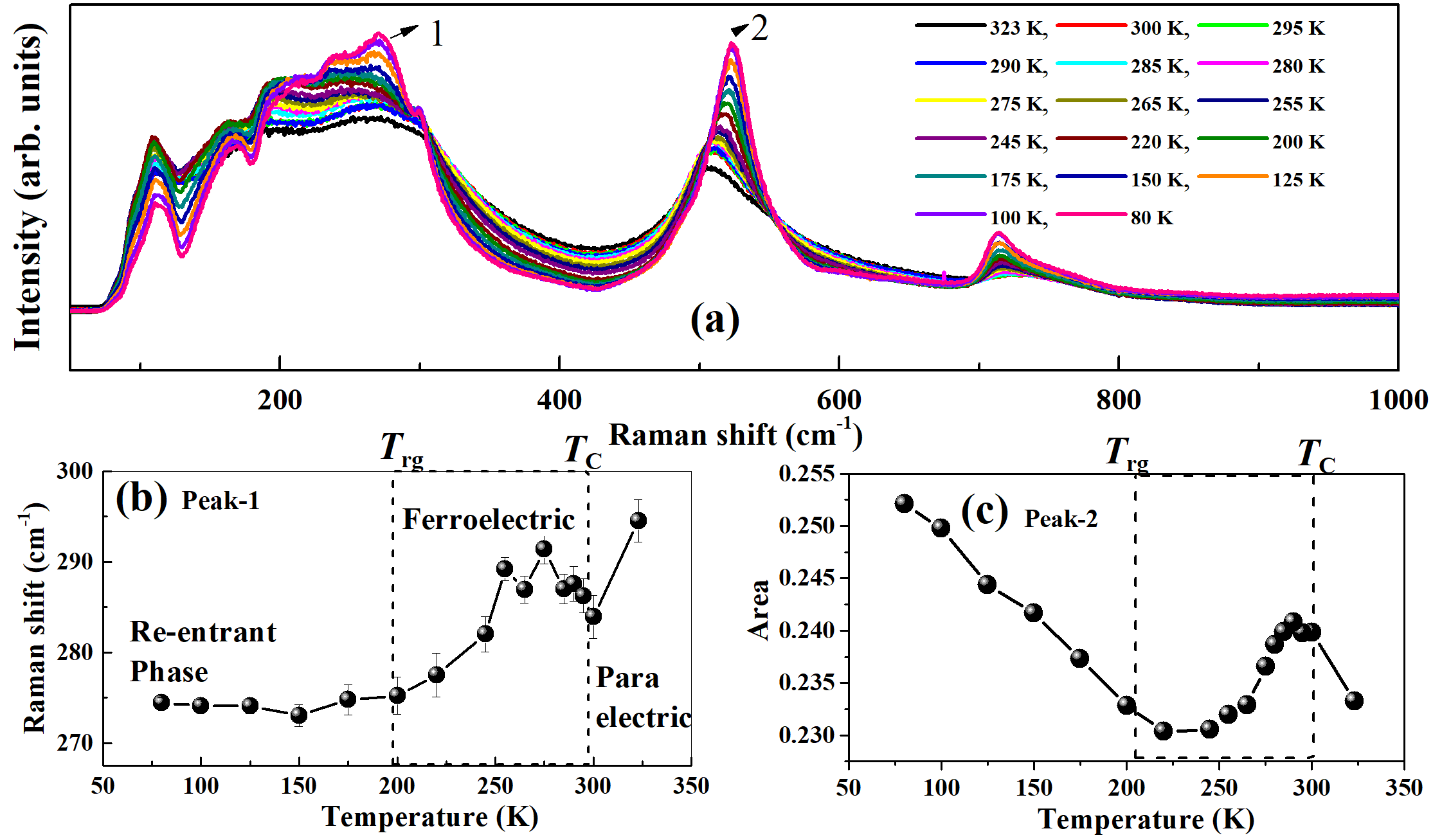}
  \caption{(a) Representative temperature dependent Raman spectra of BaSn\textsubscript{0.15}Ti\textsubscript{0.85}O\textsubscript{3}. Temperature variation of (b) Raman shift of peak-1 and (c) normalized area of peak-2. }
  \label{fig:Raman}
\end{figure}

Room temperature Raman spectroscopy as shown in Fig.~\ref{fig:Raman} shows modes resembling of rhombohedral/orthorhombic/ tetragonal modes which are symmetry forbidden for the Pm$\overline{3}$m space group, as evidenced from XRD of the same sample (Fig.~\ref{fig:xrd}). This is a consequence of the local structure deviating from the average structure which is common in materials exhibiting DPT \cite{surampalli2019evidence}. However, due to broad overlapping modes it is nearly impossible to distinguish the exact local structure near room temperature. It can be still seen from the spectra taken at 323 K ($>$ $T$\textsubscript{C}) modes still persist indicating local structure is preserved well above $T$\textsubscript{C}. This is generally observed in DPT where it is suspected that nucleation of low temperature phase starts appearing at very high temperatures ($>$ $T$\textsubscript{C}). 

Albeit abrupt changes in wave number is seen around this temperature range ($T$\textsubscript{C}), more temperature points above Tc are needed to conclude the phase transition as observed from Raman spectroscopy. It is expected in DPT the changes in local structure associated with phase transition are reflected at higher temperatures than the $T$\textsubscript{C} observed from dielectric measurements. However, this paper focuses on explaining the low temperature peculiar behavior as observed from dielectric and electric polarization data, hence, the spectroscopy analysis is limited to the temperatures below $T$\textsubscript{C}. 

Temperature dependent evolution of A(TO) mode wave number corresponding to octahedral vibrations against barium atom is show in Fig.~\ref{fig:Raman} (b). This mode is susceptible to phase transitions in pure BTO showing discontinuous jump across different phase transitions viz., Cubic - Tetragonal - Orthorhombic - Rhombohedral \cite{perry1965temperature}. The wave number of this mode shows a sharp decrease on lowering temperatures near phase transitions. Similarly, a decrease in wave number is observed in the present work, albeit gradually as a consequence of diffuse nature of phase transition. These results suggest that system transforms to rhombohedral \cite{upadhyay}  in a wide temperature range as expected for DPT systems.  Further, area under the polar TO mode is calculated ($\approx$535 cm$^{-1}$) and it is seen to be decreasing below $T$\textsubscript{C} and increasing towards low temperatures. All the spectra are corrected with bose-einstein factor before calculating the area of the TO mode, to eliminate the varying population contribution with temperature\cite{hart1970temperature}. It is well known about the structural inhomegenity in the compounds showing DPT and thus, this decrease can be attributed to the system transforming to the much more ordered state by the interplay between the increasing ferroelectric and decreasing paraelectric nature present in the system even below $T$\textsubscript{C}. These results along with wave-number trend with temperature emphasizes the DPT nature present in the system. However, an unusual increase in area is observed towards low temperatures ($<$ $T$\textsubscript{rg}). Such increase might likely be due to the increase in  structural disorder, the reason which will be discussed in the further sections. M$\ddot{o}$ssbauer technique is further employed to complement these results and are discussed in the next section.

\subsection{M$\ddot{o}$ssbauer analysis}

\begin{figure}[htbp]
  \centering
  \includegraphics[width=8cm]{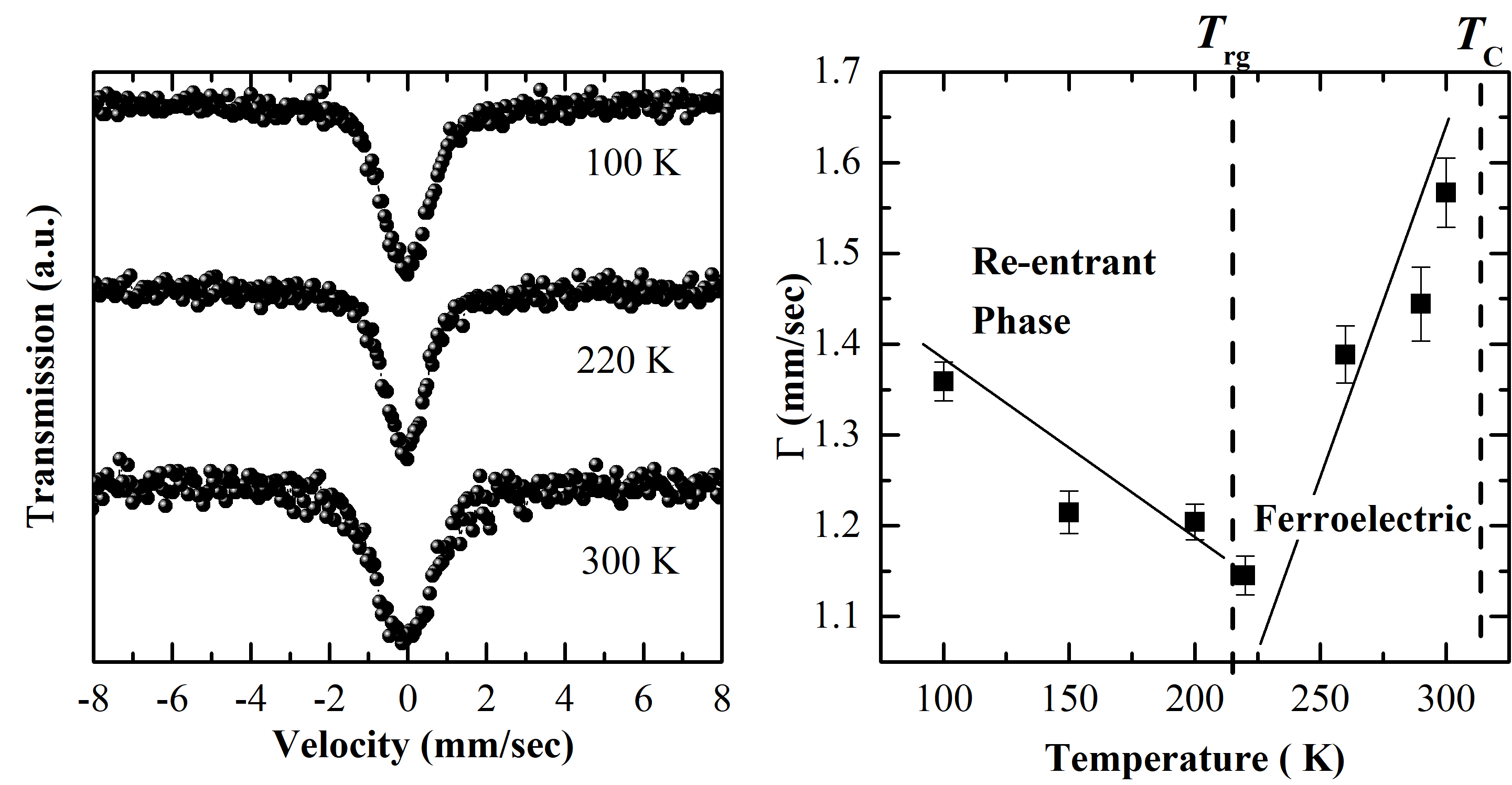}
  \caption{(Left) Representative temperature dependent \textsuperscript{119}Sn M$\ddot{o}$ssbauer spectra of BaSn\textsubscript{0.15}Ti\textsubscript{0.85}O\textsubscript{3}. (Right) Temperature variation of width of \textsuperscript{119}Sn M$\ddot{o}$ssbauer, considered to be proportional to the electric field gradient.}
  \label{fig:Mossbauer}
\end{figure}

M$\ddot{o}$ssbauer spectra taken at different temperatures between 100 - 300 K are shown in Fig.~\ref{fig:Mossbauer}. The width obtained from the spectral fitting against the temperature is also shown in the same figure. The M$\ddot{o}$ssbauer spectral width which is proportional to electric field gradient gives information about the local environment around Sn nucleus. The usual trend in ferroelectrics follow an increasing value of width below Curie temperature as a consequence of polar ordering giving rise to a non-symmetrical environment around Sn nucleus. However, below $T$\textsubscript{C} an anomalous trend is observed here where decreasing width is observed on lowering temperature contrary to the observed increase in ferroelectric transitions \cite{upadhyay2014electro}. The reason for such observation is similar to the one discussed in the earlier section and can be explained within the context the interplay between paraelectric \& ferroelectric relative phase fractions, would result in such decrease in line width below $T$\textsubscript{C}. On further lowering the temperature an increasing trend in width is observed, below $T$\textsubscript{rg}, complementing again with Raman results viz., an onset of disorder in the system. Such increase in width is also resembles to the increasing width across the relaxor transition in relaxors \cite{surampalli2019evidence}, thereby correlating the dispersion trend observed in dielectric.
 
\begin{figure}[t]
  \centering
  \includegraphics[width=8cm]{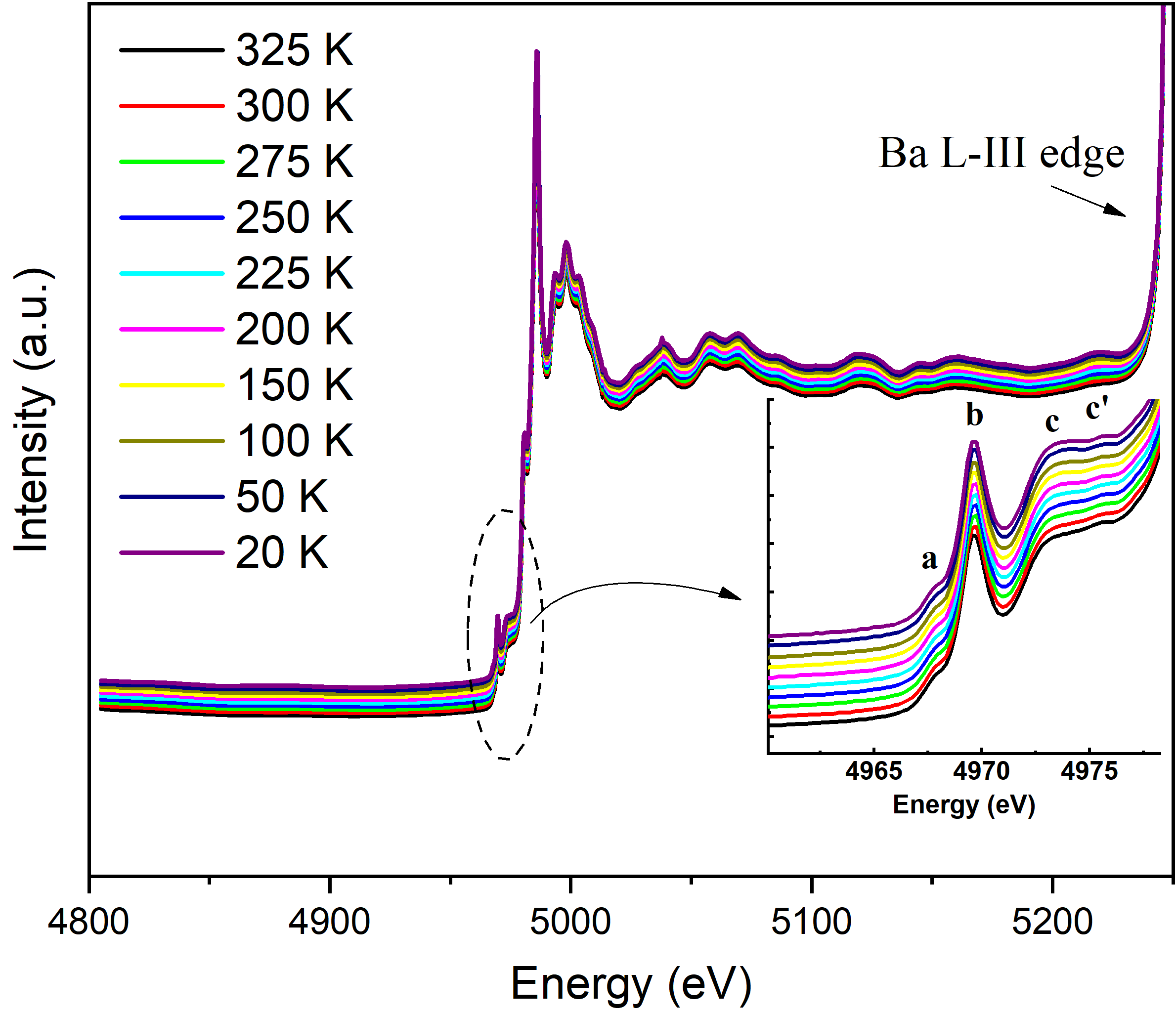}
	\includegraphics[width=5cm]{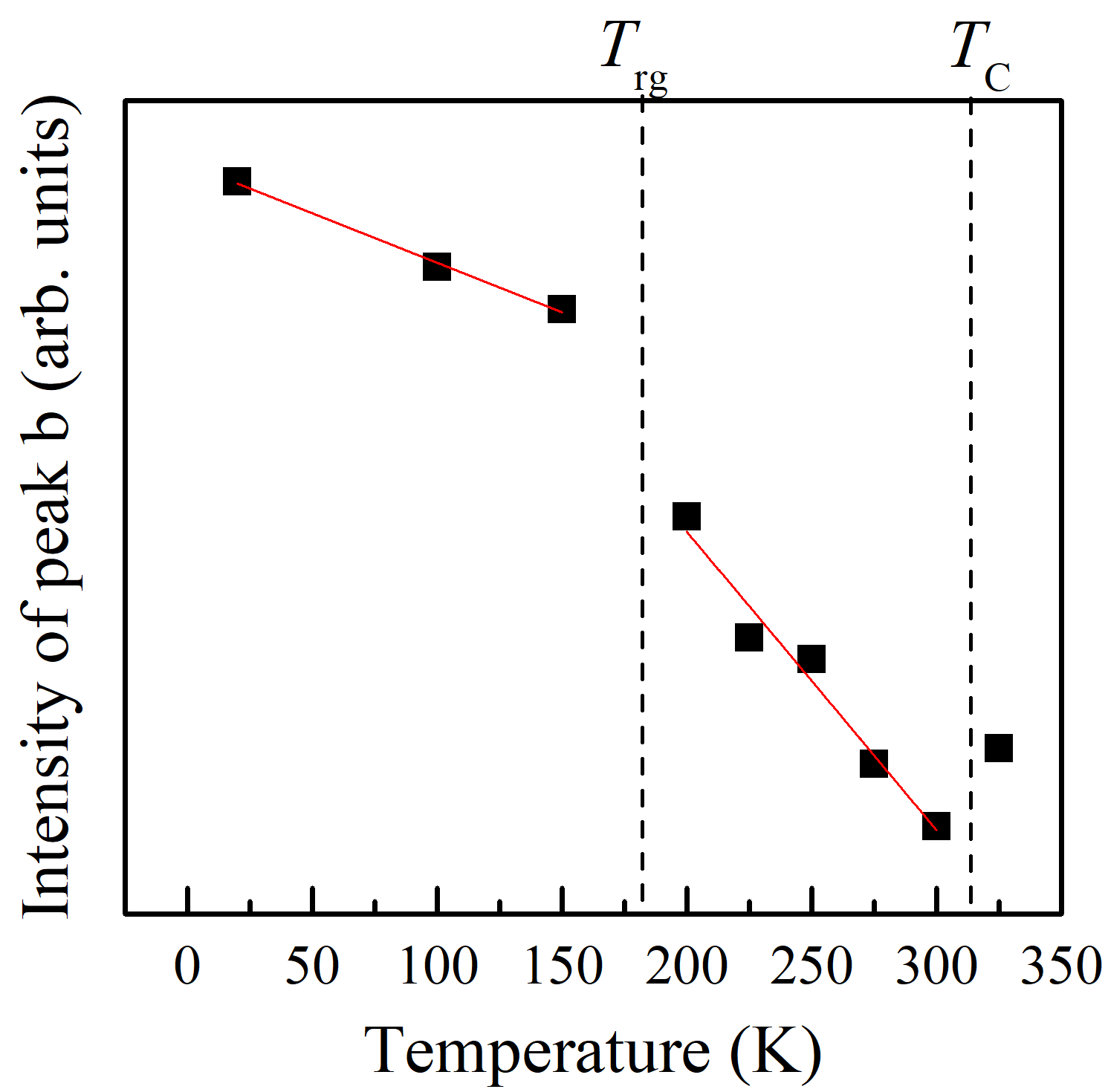}
  \caption{ (Top) Representative temperature dependent x-ray absorption spectra (XAS) of BaSn\textsubscript{0.15}Ti\textsubscript{0.85}O\textsubscript{3}. The data is vertically shifted. Inset shows the representative fitting of different modes in pre-edge region for 300 K data (Bottom) Variation of integral area of peak-b with temperature.}
  \label{fig:XANES_Results}
  \label{fig:XANES} 
\end{figure}

\subsection{Ti K- edge x-ray absorption spectroscopy investigation}

X-ray Near Edge Spectroscopy (XANES) performed near Ti- K edge at various temperatures is presented in Fig.~\ref{fig:XANES}. The inset shows pre-edge region with the peaks labeled as a, b, c, c’. The details corresponding to the origin of these peaks is discussed in literature widely \cite{surampalli2019evidence}. In particular the behavior of peak b, its intensity  with temperature reveals information about TiO\textsubscript{6} distortions. In fact it corresponds to electronic excitation to empty Ti \textit{d} states below the continuum, having quadrupolar transition rule. The \textit{p}-\textit{d} hybridization with oxygen levels enhances the dipole allowed transition probability. Such hybridization is forbidden in regular TiO\textsubscript{6} octahedron but increases for distorted octahedra and/or off center Ti displacement\cite{itie2007high}. And it is demonstrated that across the phase transitions in BTO, the octahedra are susceptible to distortions, reflecting in the changes in peak position and intensity of peak b\cite{ravel1998local}. The pre-edge features were analysed as a combination of gaussian peaks. The intensity of the peak b is plotted against the temperature in the Fig.~\ref{fig:XANES_Results} and it is evident that two different regions exist below $T$\textsubscript{C}, separated by dashed lines in the Fig.~\ref{fig:XANES_Results}, showing different behavior in intensity of peak b with temperature. In particular lowering below $T$\textsubscript{rg} the peak b raises, pointing out the raising of TiO\textsubscript{6} octahedra distortions, such off center Ti displacement allowing establishing the FE order. Below $T$\textsubscript{rg} is evident a different slope of b peak with respect to temperature change suggesting a different behaviour that may originate from  establishing local inhomogeneities related to polar nanoregions in the relaxor phase.

\maketitle \section{Summary and Conclusions}

In summary, we studied the diffuse phase transition occurring in BaSn\textsubscript{0.15}Ti\textsubscript{0.85}O\textsubscript{3}. The dielectric data confirms the system's transformation of paraelectric-ferroelectric over a broad range of temperature, a characteristic of DPT. Within this region spatial inhomogeneities in structure viz., co-existing in different phases, and is confirmed from the trends in M$\ddot{o}$ssbauer and Raman spectral analysis. The remanent polarization obtained from P-E hysteresis loops, is seen increasing on cooling till $T$\textsubscript{rg}, correlating with the trend observed in ferroelectrics. Below $T$\textsubscript{rg}, the remanent polarization starts to decrease concomitantly a significant dielectric dispersion becomes evident. This low temperature phase ($<$ $T$\textsubscript{rg}) is denoted as re-entrant phase with relaxor like properties matching with the literature viz., Vogel-Fulcher like dispersion in dielectric loss, decrease in remenant polarization and the trend in maximum polarization with electric field. The increasing line widths \& area of Mossbauer spectra and Raman spectroscopy at low temperatures on cooling further signals a much disordered state at low temperatures. 

Further, the polarization evolution with temperature is studied from FORC distribution. The FORC distribution reveals clusters with ferroelectric ordering start to form within the range of Curie temperatures. A similar phenomenon is described by Davies et.al., \citep{davies2005magnetization}, where they call this an ``isolated clusters phase".  If really isolated, they produce a FORC signature consisting of a vertical ridge along the left edge of the FORC space, entirely confined in the lower quadrant \cite{lanci2018forward}. In reality, we see a ridge that extends symmetrically in the upper and lower quadrant, and this can only be explained by the presence of a random bias field originating from interactions between these regions. So, these are definitively not isolated regions. Also, unlike in Ref\cite{davies2005magnetization}, ordering starts in a blocking state, as shown by their FORC diagram, while we start in an unblocked state, where the polarization of ordered regions decay in time. The room-temperature FORC diagram thus shows that ordering does not start at the level of single, totally isolated crystals, as one might expect from compositional variations. Instead, a sort of loose network with FE ordering is forming (maybe along grain boundaries). 

The FORC studies do not reveal any thermodynamic phase transition near $T$\textsubscript{rg}. Instead, it is seen that ferroelectric exchange field that represents the interaction strength between different regions in the sample is increasing with temperature. This behavior suggests no breaking of long-range order in the system at low temperatures and ferroelectric nature is increasing at all temperature. On the other, the regions that are susceptible to fluctuate with electric field are observed at all measured temperatures viz., relaxor like entities that exist at all temperatures. 

In conclusion the FORC analysis provides a picture of polar regions those switch like expected from an ideal single-domain behavior, but the switching field depends on time : the more time is spent at a certain field, the more likely some particles will be switched. Contrary to many magnetic nanoparticle systems, the regions are affected by exchange coupling, probably across the grain boundaries. Cooling below 100 K is expected to progressively block the entire specimen volume and produces the decrease of tan$\delta$ seen in Fig. 2. A very noticeable result is that the ratio between saturation remanent polarization and saturation polarization approaches 0.5 if calculated from desheared data, that is, as if the regions would be uncoupled. This ratio is expected for randomly oriented domains with uniaxial anisotropy. Such randomness of anisotropies in the system might result either from the micro structure of the polycrystalline or from having to contain different polar orders in the system. The latter reason might explain the increasing line widths of Raman and Mossbauer spectra, XANES anomalies at low temperatures,  as this can be explained by different polar orders are percolating throughout out the measured temperature range. The work done by Fang et.al., \cite{fang2018re} on barium titanate based disordered systems comes close to our results in which they demonstrated visually re-entrant glass as co-existing orthorombhic \& tetragonal domain structures. This might explain the existence of randomly oriented domains of uni-axial anisotropy. Hence, re-entrant glass is unlike a spin glass system, but a heterogeneous system that behaves as a glass (i.e., the disorder occurs at a much larger scale). Finally, we want to point that since the low temperature relaxor like behavior is explained within ferroelectric exchange model and thus investigating proper relaxor compositions \& other established re-entrant materials in future employing FORC studies provides basis for understanding such complex materials. \\

\maketitle \section{Acknowledgments}

AS and VRR thank Department of Science and Technology (DST) and IISc, Bangalore for travel support for performing experiments at XAFS beamline, ELETTRA. Authors thank Dr. V. G. Sathe for the Raman spectroscopy measurements. Thanks are due to Dr. Giuliana Aquilanti and Dr.Luca Olivi for the help during XAFS measurements at Elettra.  Mr. Anil Gome is acknowledged for the help with M$\ddot{o}$ssbauer spectroscopy measurements.


\bibliography{References}
 
\end{document}